\documentclass[aps,twocolumn,pra,showpacs,floatfix]{revtex4}
\usepackage{epsfig}
\usepackage{graphicx}
\usepackage{dcolumn}

\begin{document}


\title{Many-body calculations of relativistic energy shifts
for single- and double-valence atoms
}

\author{I. M. Savukov}
\email{isavukov@lanl.gov}
\affiliation{Los Alamos National Laboratory, New Mexico, 87545, USA}

\author{V. A. Dzuba}
\email{V.Dzuba@unsw.edu.au}
\affiliation{School of Physics, University of New South Wales,
Sydney 2052, Australia}

\date{\today}

\begin{abstract}

Relativistic Hartree-Fock method together with many-body perturbation
theory and configuration interaction techniques are used to calculate
relativistic energy shifts for frequencies of the strong electric
dipole transitions of C~III, C~IV, Na~I, Mg~I, Mg~II, Al~II, Al~III,
Si~IV, Ca~II and Zn~II. These transitions are used for search of the
variation of the fine structure constant in quasar absorption spectra.
The results are in good agreement with previous calculations. The
analysis of Breit contributions is also presented.

\end{abstract}


\maketitle

\section{Introduction}

Search for variation of fundamental constants is currently an
extensive area of research which spans the whole lifetime of the
Universe from Big Bang nuclear synthesis to present day atomic-clock
experiments (see, e.g. a review by Uzan~\cite{Uzan}). This search is
motivated by theories unifying gravity with other interactions as
well as by many cosmological models. In both cases a possibility for
fundamental constants to very in space and/or time is present.

Strong evidence
that the fine structure constant, $\alpha$ might be smaller about ten
billion years ago was found in the quasar absorption
spectra~\cite{Webb99,Webb01,Murphy01a,Murphy01b,Murphy01c,Murphy01d}.
This result was obtained from the analysis of the data from Keck
telescope in Hawaii by the group of researchers based at University
of New South Wales in Australia. However, the analysis of the data from VLT
telescope in Chile performed by different groups \cite{vlt1,vlt2}
gives null result. There is an outgoing debate in the literature
about possible reasons for this disagreement.

All the analysis in Refs.~
\cite{Webb99,Webb01,Murphy01a,Murphy01b,Murphy01c,Murphy01d,vlt1,vlt2}
was performed with the use of the so-called many-multiplet (MM)
method which was first suggested in Refs.~\cite{Webb99,Dzuba99}.
This method uses frequencies of strong atomic electric dipole
transitions for the analysis. Its sensitivity to variation of the
fine structure constant is more than an order of magnitude better
than the analysis of the fine structure intervals which was used
before~\cite{Wolfe,Cowie,Varsh}. This dramatic gain in sensitivity
comes with some complications. The method relies on atomic
calculations to reveal dependence of atomic frequencies on the fine
structure constant. All calculations used in the analysis so far
were performed within a single group of researches bases at the
University of New South Wales
\cite{Dzuba99a,Dzuba01,Dzuba02,Berengut04,Berengut05,Berengut06}.
Due to importance of detecting of any variation of fundamental
constants and controversy of the results it is important that atomic
calculations be also verified by independent calculations.

A positive development in this direction is the recent independent
calculation of the relativistic energy shifts in Fe~II
ion~\cite{Porsev}. Fe~II is a single most important element for the
analysis of quasar absorption spectra. It has lines which move in
opposite direction if alpha varies, and the value of this shift is
relatively large. In principle, subject to sufficient statistics, it
alone can serve as a probe of variation of the fine structure
constant in quasar absorption spectra~\cite{Porsev}. However,
calculations for Fe~II are difficult due to large number of valence
electrons.

In present work we perform further verification of the relativistic
energy shifts in atoms of astrophysical interest by considering
atoms and ions with one and two valence electrons. In the case of
single-valence electron atoms we use Dirac-Hartree-Fock method (DHF)
and relativistic many-body perturbation theory (RMBPT). For
double-valence electron atoms we use DHF, RMBPT and configuration
interaction (CI) technique. The results on this stage of
calculations are in very good agreement with previous calculations.

We also perform the analysis of role of Breit interaction.
It turns out that Breit contribution to the energies
and relativistic energy shifts are small. However, Breit interaction
gives significant contribution to the fine structure intervals brining
them to excellent agreement with experiment. Therefore, inclusion of
Breit contributions is important for the analysis of the accuracy
of calculations.

\section{Method}
\subsection{q-factor calculations}\

The difference between frequencies in QSO spectra and in the
laboratory after taking into account the Doppler shift  depends on
values of fine-structure constant $\alpha$. For small changes of
$\alpha$, the transition frequency changes linearly with
$(\alpha/\alpha_0)^2$ and can be presented in a form
\begin{equation}
  \omega(x) = \omega_0 + qx,
\label{omega}
\end{equation}
where $\omega_0$ is the laboratory value of the frequency and
$x = (\alpha/\alpha_0)^2-1$, $q$ is the sensitivity coefficient
 to be found from atomic calculations.
Note that
\begin{equation}
q=\frac{d\omega}{dx} \ \ {\rm at} \ \ x=0.
\end{equation}
To obtain q-factors theoretically, it is
necessary to calculate energies for at least two different values of
the fine structure constant. Symmetric formula for the derivative is
more accurate, and better precisions of $q$ can be reached by
calculating energies at two values of alpha symmetrically displaced
from the standard value. Due to numerical accuracy issues and
non-linear dependence in a large range, the displacement should be
not very small and not very large. We find that $\alpha_1=1/134.0$
and $\alpha_2=1/140.0$ satisfy both conditions.
The $q$ values are calculated using the following equation
\begin{equation}
q=[E(\alpha_1)-E(\alpha_2)]\alpha_0^2/(\alpha_1^2-\alpha_2^2),
\end{equation}
The energies and $q$-factors are both in inverse cm units.

\subsection{DHF and 2nd-order RMBPT}

For $q$-factors of univalent atoms and ions, we calculate energies
using the 2nd-order RMBPT formalism built on expansion in the
frozen-core $V^{(N-1)}$ DHF basis. By comparison of q-factors from
the 2nd-order and DHF calculations, we find that the precision of
second-order theory is expected to be sufficient. The 2nd-order
RMBPT formalism is described in Ref.\cite{secmbpt}. In the DHF basis
first-order correction is already included, and the number of
second-order corrections is reduced.  The summation over excited
states in the second-order expressions is carried out by using a
finite compact B-spline DHF basis, with cavity sizes chosen to
minimize influence of boundary conditions on the valence energies of
interest. Angular orbital momentum of excited states is limited to
5, without much reduction in the precision. Other parameters are
chosen to minimize numerical errors.

The second-order RMBPT gives much more accurate energies than the
DHF theory as can be seen from our calculations presented in
Table~\ref{table1}. We also compare theoretical and experimental
fine-structure splitting between $p_{3/2}$ and $p_{1/2}$ states. The
agreement is very good in the second-order of RMBPT. However, it is
further significantly improved when the Breit corrections discussed
in next section are added. This is another indication of  high
accuracy of the calculations.

\begin{table}
\caption{DHF and 2nd-order energies and fine-structure splittings of
univalent atoms/ions} \label{table1}
\begin{tabular}{lrrrrrrr}
\hline
\multicolumn{1}{c}{Element} & \multicolumn{1}{c}{State} &
\multicolumn{3}{c}{Energy} &
\multicolumn{3}{c}{Fine Structure}\\
 & &\multicolumn{1}{c}{DHF} &\multicolumn{1}{c}{E2\footnotemark[1]} &
\multicolumn{1}{c}{Expt\footnotemark[2]} &
\multicolumn{1}{c}{E2\footnotemark[1]} &
\multicolumn{1}{c}{E2+Br\footnotemark[3]} &
\multicolumn{1}{c}{Expt\footnotemark[2]} \\
\hline \hline
 C IV   & 2p$_{1/2}$ &  65201 &  64548 &  64484 &      &      &      \\
        & 2p$_{3/2}$ &  65328 &  64680 &  64592 &  132 &  107 &  108 \\
 Na I   & 3p$_{1/2}$ &  15921 &  16812 &  16956 &      &      &      \\
        & 3p$_{3/2}$ &  15937 &  16831 &  16973 &   18 &   17 &   17 \\
        & 4p$_{1/2}$ &  28904 &  30068 &  30267 &      &      &      \\
        & 4p$_{3/2}$ &  28909 &  30074 &  30273 &    6 &    6 &    6 \\
 Mg II  & 3p$_{1/2}$ &  34530 &  35603 &  35669 &      &      &      \\
        & 3p$_{3/2}$ &  34620 &  35700 &  35761 &   97 &   91 &   92 \\
        & 4p$_{1/2}$ &  78574 &  80463 &  80620 &      &      &      \\
        & 4p$_{3/2}$ &  78605 &  80496 &  80650 &   32 &   30 &   31 \\
 Al III & 3p$_{1/2}$ &  52709 &  53672 &  53683 &      &      &      \\
        & 3p$_{3/2}$ &  52944 &  53919 &  53917 &  247 &  233 &  234 \\
        & 4p$_{1/2}$ & 141252 & 143538 & 143633 &      &      &      \\
        & 4p$_{3/2}$ & 141334 & 143623 & 143714 &   85 &   80 &   80 \\
 Si IV  & 3p$_{1/2}$ &  70540 &  71309 &  71288 &      &      &      \\
        & 3p$_{3/2}$ &  71008 &  71794 &  71749 &  486 &  461 &  461 \\
        & 4p$_{1/2}$ & 215704 & 218226 & 218267 &      &      &      \\
        & 4p$_{3/2}$ & 215870 & 218397 & 218429 &  171 &  162 &  162 \\
 Ca II  & 4p$_{1/2}$ &  23403 &  25490 &  25192 &      &      &      \\
        & 4p$_{3/2}$ &  23603 &  25722 &  25414 &  232 &  225 &  223 \\
 Zn II  & 4p$_{1/2}$ &  44610 &  48548 &  48481 &      &      &      \\
        & 4p$_{3/2}$ &  45347 &  49429 &  49355 &  881 &  866 &  874 \\
\hline
\end{tabular}
\noindent \footnotetext[1]{DHF+2nd-order}
\noindent \footnotetext[2]{NIST, Ref.~\cite{NIST}}
\noindent \footnotetext[3]{E2+Breit (see Table~\ref{Table2})}
\end{table}

\subsection{Breit corrections}

Relativistic energy shift which was considered above is due to
the difference between Dirac and Schr\"{o}dinger equations.
This difference leads to a correction to the energy proportional
to $\alpha^2$ in the leading order. Therefore, for small change
of $\alpha$ this correction coincides with the definition
of the $q$-coefficient (see formula (\ref{omega}).
However, there is also Breit relativistic correction to the
inter-electron interaction~\cite{Breit}. This correction is
also proportional to $\alpha^2$ and therefore contributes
to the $q$-coefficients. It is important to check the values
of these corrections to have reliable results.
We include Breit interaction using the technique developed in our
previous works~\cite{Breit1,Breit2}.

We use the following form of the Breit operator
(atomic units)
\begin{equation}
    \hat H^B = - \frac{{\hat{\bf \alpha}_1}\cdot{\hat{\bf \alpha}_2}+
    ({\hat{\bf \alpha}_1}\cdot{\bf \hat{n}})({\hat{\bf \alpha}_2}
    \cdot{\bf \hat{n}})}{2r}.
\label{HBreit}
\end{equation}
Here ${\bf r} = {\bf \hat{n}}r$, $r$ is distance between electrons and
$\hat{\bf \alpha}_i$ is the $\alpha$-matrix of the corresponding electron.
This is a low frequency limit of the relativistic correction to the Coulomb
interaction between electrons. It contains magnetic interaction and
retardation.

Similar to Coulomb interaction, Breit interaction creates a
potential which is to be added to Hartree-Fock potential.
In the case of closed-shell atoms  direct term in Breit
potential vanishes and only exchange term remains.
This is the case for single-valence-electron atoms
considered in present work since we use the $V^{N-1}$ approximation.

Self-consistent calculations are performed for a closed-shell
core in a potential which is a sum of Coulomb and Breit terms
\begin{equation}
  \hat V = \hat V^C + \hat V^B.
\label{VCB}
\end{equation}
States of valence electrons are calculated in the same potential
(\ref{VCB}). In this approach Breit interaction between electrons
receives exactly the same treatment as the Coulomb one. It is
first included as interaction between core electrons and then as
an interaction between valence and core electrons. Therefore, an
important effect of core relaxation is included. A less important
effect of Breit interaction on inter-electron correlations is not
included in present work. This is justified by small value of
the corrections.

Note that non-perturbative treatment of Breit interaction leads
to inclusion of higher-order in Breit operator
terms, terms proportional to $(\hat H^B)^2$,$(\hat H^B)^3$, etc.
Inclusion of these terms cannot be justified and in principle
they can be easily eliminated by a rescaling
procedure in which Breit operator is suppressed in the
calculations by a scaling parameter $\lambda$ and then the
answer is interpolated to $\lambda=1$.
It turns out, however, that as a rule $\lambda=1$
is already in linear regime.

The values of the corrections are
found by running programs with and without corresponding extra
terms in the potential. The calculated values are presented in
Table~\ref{Table2}. In Table~\ref{table1} we also included Breit
corrections to the fine structure splitting.

Calculations show that although first-order valence Breit correction
is a dominant contribution among Breit corrections to the energy of
valence electrons, transition energy has a substantial cancelation
for this correction, and Breit core-relaxation contribution becomes
comparable with the valence Breit contribution for the transition
energies and hence for q-values. The inclusion of core-relaxation
effect significantly changed the value of Breit contribution.

\subsection{BO+CI method}

To calculate alpha variation coefficients for divalent atoms and
ions, we will use the Brueckner-orbital (BO)+CI method, introduced
in Ref.\cite{BOCI}, and modified in Ref.\cite{CIMBPTBr} to include
first-order Breit corrections. This method is essentially the
combination of CI, to treat strong valence-valence interactions, and
MBPT, to treat important valence-core interactions. It is similar to
the method discussed in Refs.\cite{CI+MBPT1,CI+MBPT2,CIBS}.

The BO-CI method described in detail in Ref.\cite{BOCI} is based on
the effective Hamiltonian formalism which leads to the problem of
diagonalization of the Hamiltonian matrix built on the two-electron
configuration state functions. Beyond the frozen-core Hamiltonian
the first-order electron-electron interaction Hamiltonian and
second-order correction which consists of the two-particle screening
correction and the one-particle self-energy correction are included.
In the BO-CI method, the basis functions are chosen as BO and
include second-order self-energy corrections together with DHF
potential. The residual two-particle Hamiltonian matrix, that
includes first-order valence-valence interaction and second-order
Coulomb screening interaction, is evaluated in the BO basis and
diagonalized to obtain state energies and CI wave functions.

\section{Results}
\begin{table}
\caption{Calculations of $q$-factors for univalent atoms/ions}
\label{Table2}
\begin{tabular}{lccccc}
\hline \hline
Atom/Ion & State & DHF+2nd & Breit& Total & Other\footnotemark[1] \\
\hline
C IV &  3p1/2   &    102 &   13   &     115   & 104(20) \\
     &  3p3/2   &    233 &  -12   &     221   & 232(20) \\
Na I & 3p1/2    &     45 &   -1   &      44   & 45(4) \\
     & 3p3/2    &     63 &   -2   &      61   & 63(4) \\
     & 4p3/2    &     59 &   -2   &      57   & 59(4) \\
     & 4p1/2    &     53 &   -2   &      51   & 53(4) \\
Mg II& 3p1/2    &    119 &    1   &     120   &120(10) \\
     & 3p3/2    &    216 &   -5   &     211   &211(10) \\
     & 4p1/2    &    167 &   -6   &     161   &        \\
     & 4p3/2    &    200 &   -8   &     192   &        \\
Al III&  3p1/2  &    218 &    5   &     223   & 216(14) \\
     &  3p3/2   &    466 &   -9   &     457   & 464(30) \\
     &  4p1/2   &    349 &  -12   &     337   &        \\
     &  4p3/2   &    434 &  -17   &     417   &        \\
Si IV&  3p1/2   &    347 &   13   &     360   & 346    \\
     &  3p3/2   &    835 &  -12   &     823   & 862    \\
     &  4p1/2   &    617 &  -20   &     597   &       \\
     &  4p3/2   &    789 &  -29   &     760   &      \\
Ca II&  4p1/2   &    219 &    3   &     222   &    224 \\
     &  4p3/2   &    454 &   -4   &     450   &    452 \\
Zn II&  4p1/2   &   1590 &   -5   &    1585   &   1584(25) \\
     &  4p3/2   &   2508 &  -20   &    2488   &   2479(25) \\
\hline
\end{tabular}
\noindent \footnotetext[1]{Ref.~\cite{archDzuba}}
\end{table}

\begin{table}
\caption{Calculations of $q$-factors for divalent atoms/ions}
\label{Table3}
\begin{tabular}{lccccc}
\hline \hline
Atom/Ion & State & CI+MBPT & Breit& Total & Other\footnotemark[1] \\
\hline
C III &  2s2p &           &      &   163 &       165 \\
Mg I  &  3s3p &         93&   -7 &    85 &       86(10)\\
      &  3s4p &         89&   -8 &    80 &       87  \\
Al II &  3s3p &           &      &   270 &      270(30) \\
\hline
\end{tabular}
\noindent \footnotetext[1]{Ref.~\cite{archDzuba}}
\end{table}

Results of calculations of $q$-factors for univalent and divalent
atoms and ions are presented in Tables \ref{Table2} and
\ref{Table3}, respectively. Accurate agreement is achieved between
our second-order values and $q$-factors previously reported and
compiled in Ref.\cite{archDzuba}. Univalent atoms and ions are
calculated from second-order RMBPT energies with the method
described in the previous section. Because the contribution from the
second order turned out to be relatively small, we expect that
second-order results will give quite reliable values. Because
previously only dominant relativistic effects were included within
DHF formalism, we also added Breit  corrections to energies
to investigate the effects beyond Dirac-Fock approximation.

For divalent atoms and ions calculations are performed with the
BO+CI code, which is described in the previous section. All
first-order Breit corrections introduced into CI+MPBT in
Ref.\cite{CIMBPTBr} are also included.

\section{Conclusions}

We have calculated $q$-factors for mono- and divalent atoms and ions
of interest for the extraction of fine-structure variation from
quasar spectra. Our results agree with good accuracy with previous
calculations and provide necessary independent verification. In
particular, more difficult for theory divalent atoms and ions are
calculated with a new method, BO+CI. Breit  corrections, ignored
previously, have been also evaluated. Although found in this work to
be small, potentially they constitute a dominant class of
relativistic corrections beyond the Dirac-Fock formalism.

\begin{acknowledgments}

The authors are grateful to J.S.M. Ginges for useful discussions.
The work is partly supported by the Australian Research Council.

\end{acknowledgments}

\end{document}